\documentclass{aa}  \usepackage{graphicx}
\begin{document}

   \title{Scandium: A key element for understanding Am stars}
   \author{F. LeBlanc\inst{1} and G. Alecian\inst{2}}

   \offprints{F. LeBlanc \\ \email{francis.leblanc@umoncton.ca}}

   \institute{ D\'epartement de Physique et d'Astronomie,
               Universit\'e de Moncton,
               Moncton, NB, E1A 3E9, Canada
               \and
               LUTH (Observatoire de Paris - CNRS), 
               Observatoire de Meudon, 92195 Meudon Cedex, France
             }

   \date{Received {\it date will be inserted by the editor}\
         Accepted {\it date will be inserted by the editor}
        }

\abstract{ {\rm Context.} Atomic diffusion is believed to cause
the abundance anomalies observed in AmFm stars. However, the detailed process has
still not been well-established. For instance, two possible scenarios for the
diffusion theory are presently envisaged. They differ mainly by the depth from
which the abundance anomalies emanate. The first scenario predicts that the
abundances are modified in the superficial regions of the star, just below the
hydrogen convection zone. The second scenario predicts that a much deeper
extension of the mixing zone exists due to the convection caused by Fe accumulation
in regions below the hydrogen convection zone.\\ {\rm Aims.}  We
calculate much more accurate radiative accelerations of Sc than previously,
to better understand the observed abundance anomalies of this element. We
believe that it is a key element to use as a diagnostic tool for
understanding AmFm stars. \\ {\rm Methods.} The method employed to obtain these
radiative accelerations is based on an interpolation from the parameters of
the so-called SVP parametric method. \\ {\rm Results.} The radiative
accelerations, shown here in a typical Am stellar model, are discussed in
light of the observed anomalies of Ca and Sc. Our results suggest that the
deeper mixing scenario is not entirely satisfactory: the mixing zone should be
deeper than what is predicted by recent models to account for observed Sc
underabundances. Our results seem more  compatible with the scenario where the
abundances anomalies are created in the superficial regions. However, only
detailed evolutionary modelling with mass loss and diffusion of all important
species, including Ca and Sc, with accurate radiative accelerations, will be
able to give more insight into where the source of these anomalies occur in
AmFm stars.
   }

\titlerunning{Scandium in Am stars}
\authorrunning{LeBlanc \& Alecian}

\maketitle

\section{Introduction}

Among the main observational peculiarities of Am stars are the superficial
underabundances of Ca and Sc, as well as an overabundance of iron-peak elements
(e.g. Preston 1974). The classical explanation of the AmFm star phenomenon is
that these stars, which are slow rotators, are supposed to have smaller
large-scale (mixing) motions in their envelope than \emph{normal} main sequence
stars. Since these stars are almost always in binary systems, their slow
rotation is believed to be due to tidal forces (Abt 1961). The relative
stability of their envelope allows for a large proportion of He atoms to sink
towards the deeper regions due to gravitational settling, leading to the
disappearance of the He convection zone. The abundance anomalies were then
thought to be produced by the diffusion process (Michaud 1970) at the bottom of
the hydrogen convection zone (Watson 1971, Smith 1971, Michaud et al. 1983,
Alecian 1996). At that depth, Ca is mostly in its Ar-like configuration so its
radiative acceleration is weak because of its noble gas configuration, thus
leading to its underabundance in the superficial hydrogen convection zone. The
radiative acceleration ($g_{rad}$) is due to the momentum transfer from photons
to atoms during photoabsorption. The value of $g_{rad}$ of a given element
depends on its absorption cross-section, on its abundance and on the local
physical conditions of the plasma (e.g. Alecian \& LeBlanc 2000). The diffusion
process that can create over or underabundances at various depths is a non
linear, time-dependent process that can affect the structure and the evolution of
stars.

Numerical modelling of Ca and Sc time-dependent diffusion in AmFm stars was
carried out by Alecian (1996). This modelling corresponds to the scenario first
proposed by Watson (1971), but including a bulk velocity due to mass loss (as
considered for $\lambda$ Bootis stars by Michaud \& Charland 1986).
Superficial abundances of these elements appear to depend closely on the
age of the star, on the mass loss rate, and they are very sensitive to the position of
the bottom of the superficial mixing zone (hydrogen convection zone plus
overshooting). In these computations, the stellar model was static and no
structure evolution was considered (the star was kept at the same position on
the main sequence), only diffusion was time-dependent.

Recent evolutionary models including the effect of the diffusion of the elements
show that a Fe convection zone can occur due to its accumulation at certain
depths (Richard, Michaud \& Richer 2001 and Richer, Michaud \& Turcotte 2000),
and suggest that the abundance anomalies emanate from deeper layers (at
temperatures of approximately $200\,000$ K). At that depth, a large proportion
of Ca is also in a noble gas configuration (Ne-like) which explains its observed
underabundance. However, these computions are done without mass loss. Mass loss
has two effects: it smooths the abundance inhomogenietes, and it shifts the
abundance stratifications produced by diffusion towards shallower layers in the
star (Michaud \& Charland 1986; Alecian 1996). Therefore, mass loss may affect
iron accumulation and thus the Fe convection zone predicted by these
evolutionary models.

The study of Ca diffusion is relatively easy because its atomic data are rather
well known. Indeed, a good knowledge of the atomic data is essential since
radiative accelerations are, with gravity, among the main ingredients in the
diffusion equation. The situation for Sc is quite different. For instance, both
the OPAL opacities (e.g. Rogers \& Iglesias 1992) and the Opacity Project
(hereafter OP, Seaton et al. 1992) do not include scandium in their databases.
The Kurucz (1990) data include only the first nine Sc ions. The detailed study
of Sc diffusion has never been done. Alecian (1996) calculated the $g_{rad}$ of
Sc based on the parametric method of Alecian \& Artru (1990a,b). But, as
mentioned in his study, the calculated $g_{rad}$  are not accurate enough since
extrapolations had to be done to complete the data for highly ionized Sc
ions. In this paper, we aim to calculate the $g_{rad}$ of Sc by an interpolation
method based on the parametric form of the $g_{rad}$ formulae of Alecian \&
LeBlanc (2002) and LeBlanc \& Alecian(2004) (hereafter respectively Papers I and
II).

Because scandium underabundance, along with the one of calcium, characterizes
AmFm stars, it is a key element to better understand the AmFm phenomenon.
Scandium is in an ionisation state with noble gas configuration in layers close
to the bottom of the H convection zone, and close to the bottom of the Fe
convection zone predicted by Richer, Michaud \& Turcotte (2000), but not at the same
positions as for the corresponding calcium configurations. Any model of AmFm
stars should be able to explain underabundances for both elements. Therefore,
the study of Sc diffusion could possibly shed light on which one of the two
diffusion scenarios, diffusion below the H convection zone while including mass
loss (Alecian 1996; hereafter \emph{scenario A}), and diffusion below the Fe
convection zone (Richer, Michaud \& Turcotte 2000; hereafter \emph{scenario B}),
is more likely to explain the abundance anomalies of these stars.

In Sec.~\ref{sec:grad}, we present $g_{rad}$ calculations
for Sc while using an interpolation method related to the new parametric
formulae described in Papers I and II. In Sec.~\ref{sec:abundances}, we
discuss our $g_{rad}$ with respect to the observed abundance anomalies of Ca and
Sc in AmFm stars. A short conclusion follows.

\section{Radiative acceleration calculations} \label{sec:grad}

\subsection{Method used for $g_{rad}$ calculations}

Recently, an improved parametric method for calculating $g_{rad}$ at large
optical depths was developed (Papers I and II). The improvements, as compared
to the method of Alecian \& Artru (1990a,b), were made possible because of the
use of OP opacities and data (Seaton et al. 1992). Once the parameters
of the various ions are calculated, the parametric method has the advantage of
giving relatively accurate $g_{rad}$ without having to deal with the enormous
amount of atomic data normally necessary to obtain $g_{rad}$. The parameters for
12 trace elements (C, N, O, Ne, Na, Mg, Al, Si, S, Ar, Ca and Fe) were
calculated in Paper II.

The method used here to calculate the $g_{rad}$ of Sc is based on this
parametric form of $g_{rad}$ formulae (SVP method) of both bound-bound (Alecian
1985, Alecian \& Artru 1990a, Papers I and II) and bound-free (Alecian 1994,
Papers I and II) transitions. In this method, radiative accelerations due to
atomic lines depend on 4 parameters for each ion (Eq.~(1) of Paper II),
calculated at layers where the relative population of the ion $i$ is near its
maximum.  The first parameter $\phi_{i}^{*}$ (Eqs.~(10) to (13) of Paper I) is
related to the strength of the bound-bound transitions through a weighted
average of the $gf$ values. The second parameter $\psi_{i}^{*}$ (see Eq.~(14) of
Paper I) is related to the average width of the line profiles and controls
saturation. A third parameter $\xi_{i}^{*}$ is related to the ion contribution
to the total opacity (we will neglect it for Sc since it is of little importance
for less abundant elements). Meanwhile, the value of the parameter $\alpha_i$
\footnote{This parameter characterizes an average Voigt profile of the lines for
the ion under consideration, see Eq.~(1) of Paper~II.} is determined by fitting
our parametric accelerations to those obtained by a more accurate method
(Seaton 1997, see Paper I for more details).

A similar parametric equation was also used for bound-free transitions (Eq.~(9)
of Paper II). This approximate formula is less accurate than for bound-bound
transitions and needs the knowledge of energy levels for each ion and the
computation of the partition functions at each model layer. Two other parameters
$a_{i}$ and $b_{i}$ (correction parameters for bound-free acceleration) are
determined for each ion by a same kind of fitting procedure as employed for
lines. These six parameters, along with the formulae of Paper II, define the SVP
method. They are calculated here, in an Am stellar model, for the 12 trace
elements mentioned above. These elements are treated by Seaton (1997), and our
fitting procedure is achieved using version OPCD~2.1 of the tables available at
CDS\footnote{http://vizier.u-strasbg.fr/topbase/op.html}.

The accurate evaluation of radiative accelerations necessitates detailed and
complete atomic data (including all strong transitions) for each element. This
is still lacking for scandium. A rather large data base for energy levels can be
found in the NIST Atomic Spectra
Database\footnote{http://physics.nist.gov/PhysRefData/ASD/index.html}, but this
database is incomplete for bound-bound transitions, especially in regards to
transition probabilities.

As shown in Alecian \& Artru (1990b), parameters $\phi_{i}^{*}$ and
$\psi_{i}^{*}$ generally vary quite monotonously along isoelectronic sequences.
It is then possible to approximate these parameters for some of the ions of the
elements for which atomic data are lacking, by interpolating between the nearest
isoelectronic neighbours with known parameters, to the ions in question. The
interpolation is done via the atomic charge variable. We have then considered
that to apply this interpolation method to Sc ions, using well calculated
parameters from isoelectronic neighbours, is preferable than to calculate
directly the parameters, using incomplete atomic data. Also, it would be
impossible to do the fitting procedure of Paper II, since the $g_{rad}$ of Sc
are not included in Seaton (1997) or elsewhere.

We have interpolated the
parameters for the bound-bound transitions of the ions ScIII to ScXX, but not
for ScII since parameters for an isoelectronic neighbour are missing (namely the
parameters for CaI). The parameters which were directly evaluated by Alecian \&
Artru (1990b) for ScII from atomic data were then used. Parameters $a_{i}$ and
$b_{i}$ were respectively set to 1 and 0 (this corresponds to no correction).
Relative populations of Sc ions and partition functions have been calculated
using the atomic energy level data of the NIST Atomic Spectra Database.

\subsection{Numerical results}

The results shown in this section have been obtained for a 1.9 $M_{\odot}$
stellar model with $T_{\rm{eff}} = $ 7610 K and an age of 807 Myr
(model named 1.9P1 in Talon, Richard \& Michaud 2006).

\begin{figure}[th]
\includegraphics[width=3.6in]{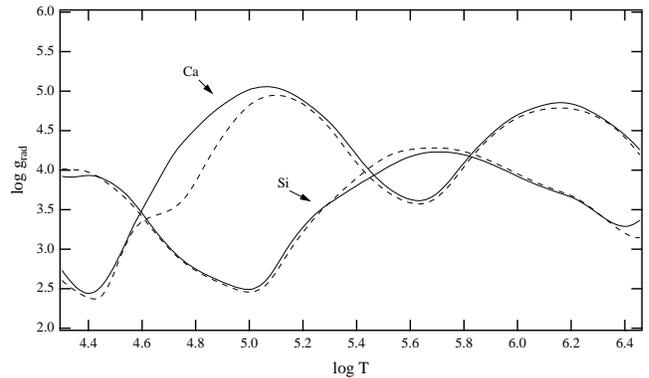} \caption{ Radiative accelerations of
Si and Ca in a model with $T_{\rm eff}=$ 7610 K (1.9 $M_{\odot}$). The solid
lines represent the results of the SVP method using known parameters, while the
dashed lines represent the $g_{rad}$ using the parameters obtained by our
interpolation method.
\label{fig:simu}
}
\end{figure}

To verify the accuracy of this interpolation method along isoelectronics
sequences, we have simulated it for elements for which the parameters are
obtained by detailed computations, and for which the SVP method is known to work
well, i.e. Si and Ca (see Paper II). Note that for these elements, accurate
radiative accelerations can also be calculated by the Seaton (1997) method,
which makes the verification process especially confident. We have compared for
these elements, accelerations obtained using their \emph{true} parameters with
those (assuming that they are unknown) obtained with parameters calculated by
interpolation along isoelectronics sequences. Results confirmed the validity of
our interpolation method, at least for Si and Ca (see Fig.~\ref{fig:simu}). The
parameters of some weakly ionized ions cannot be obtained by interpolation since
they do not have isoeletronic neighbours with known parameters. In order to
include a maximum number of ions, we have used the parameters $\phi_{i}^{*}$
and $\psi_{i}^{*}$ of SiII, CaII and CaIII from Alecian \& Artru (1990b). The
results for Ca are not very accurate at low temperatures. This is due to the
fact that the parameters $\phi_{i}^{*}$ do not, in this case, vary monotonously
for weakly ionized ions (see the figures in Alecian \& Artru, 1990b). The
interpolation method underestimates contributions of CaIV and CaV to the
acceleration. Nevertheless, deeper in the star, both methods give very similar
results. Although we only study the case of Sc in this work, we consider that
this interpolation method could be extended to other elements for which the
atomic data are not completely known, provided that their ions have
isoelectronic neighbours for which parameters are well determined.

\begin{figure}[th]
\includegraphics[width=3.6in]{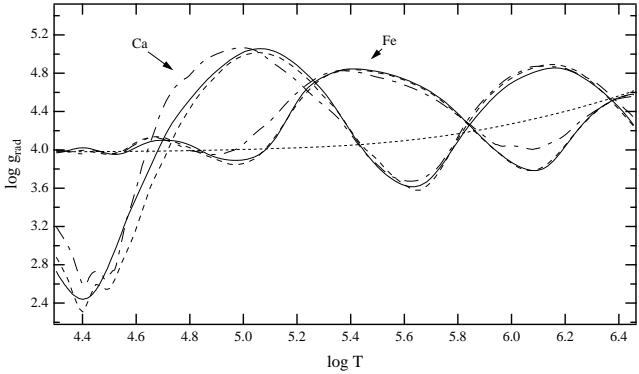}
\caption{Radiative accelerations of Ca and Fe using solar abundances calculated
with the SVP method (solid lines) and those of Seaton (1997) (dashed lines) in
the same stellar model as in Fig.~\ref{fig:simu}. The dashed-dotted lines
represent the accelerations from the models of Talon, Richard \& Michaud (2006)
using the local abundances obtained by the diffusion process. The local gravity
is represented by the dotted curve.
\label{fig:gradFe}
}
\end{figure}

\begin{figure}[th]
\includegraphics[width=3.6in]{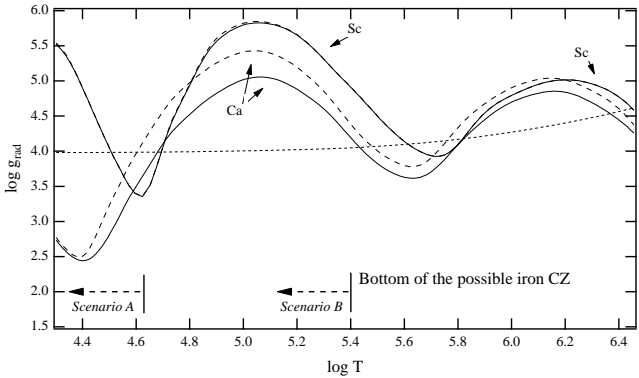}
\caption{Radiative accelerations of Ca and Sc for abundances of 1 (solid
lines) and 0.1 (dashed lines) times their solar values in the same stellar model
as in Fig.~\ref{fig:simu}. The Sc accelerations are
calculated by our interpolation method and those of Ca by the SVP method.
Also shown are the approximate locations of the bottom of the convection zones of \emph{scenarios A} and \emph{B}, and the local gravity (dotted line).
\label{fig:gradSc}
}
\end{figure}

Figure~\ref{fig:gradFe} shows the Ca and Fe radiative accelerations for a solar
abundance using the SVP method and those obtained by applying the method of
Seaton (1997)\footnote{With version OPCD~2.1 of data and codes available at
Centre de Donn\'ees de Strasbourg.}. Also shown, are the $g_{rad}$ found by
Talon, Richard \& Michaud (2006)\footnote{These accelerations were kindly
communicated by O.Richard.}. These $g_{rad}$ are calculated  via the opacity
sampling method (e.g. LeBlanc, Michaud \& Richer 2000) using the local abundance
obtained by the diffusion process. We may conclude, from these, that the results
from the SVP method are satisfactory.

Figure~\ref{fig:gradSc} shows the $g_{rad}$ of Sc (calculated by the
interpolation method) and Ca (calculated by the SVP method). The parameters for
Ca are the same as those used for the accelerations shown in
Fig.\ref{fig:gradFe}. The parameters for Sc are interpolated from those of Ca
and Fe. Two abundances are shown: 1 and 0.1 times their solar values. The two
valleys seen in the curves presented are related to the Ar-like and Ne-like
configurations, where the $g_{rad}$ are naturally weaker. We can observe that
the two curves for Sc are quite similar because this element is not very
abundant and, at the abundances used here, the $g_{rad}$ are completely
unsaturated and near their maximal values.

Generally, we consider that the SVP method has an accuracy of about 30\% (or 0.1 dex) compared to detailed methods in the temperature range of interest here (Paper II). We can suppose that the accuracy is of the same order of magnitude for our $g_{rad}$ estimation for Sc.
Unlike the curve shown for Ca in Fig.~\ref{fig:simu}, the $g_{rad}$ of Sc at
low temperatures does not seem to have abnormal behaviour. This probably means
that these $g_{rad}$ are relatively good throughout the model.
However, only
detailed $g_{rad}$ calculations could confirm this.

\section{The relative abundances of Ca and Sc in Am stars}
\label{sec:abundances}

The principal characteristic of Am stars is the presence of Ca and Sc
deficiencies at their surface. Several observational studies have evaluated the
abundances for these two elements as well as others in Am stars (e.g. Adelman
1994, Hui-Bon-Hoa, Burkhart \& Alecian 1997, Hui-Bon-Hoa \& Alecian 1998,
Hui-Bon-Hoa 1999, Griffin 2002). These studies show that generally,
Sc is more underabundant (relative to solar values) than Ca in Am stars.

When neglecting all other factors than
gravity and radiative acceleration, if $g_{rad} > g$ the element will be driven upwards.
While if $g_{rad} < g$ the element will migrate towards the centre of the
star. At first approximation, since  $g_{rad} > g$ means also that a larger abundance of the element can be supported by the radiation field, one considers that the element should become overabundant (with respect to solar value) in those layers, and underabundant in layers where $g_{rad} < g$.
Of course, time-dependent diffusion, which is a non-linear process, is more
complicated than what is described by this simplified approach.

As mentioned previously, two possible scenarios are suggested to explain these
underabundances which both depend on the diffusion process.

\emph{Scenario A} supposes that the surface abundances emanate from the regions
below the superficial hydrogen convection zone, assuming a global mass flow
(defined by the mass loss rate). These regions are near where the Ar-like
configurations of Sc and Ca dominate, thus explaining their underabundance. 
Alecian (1986, 1996) showed that models with
diffusion while including mass loss, can mimic a deepening of the surface mixing
zone, and results in varying superficial abundances as a function of
time. This study showed that the surface abundance of Ca and Sc for Am
stars depend on both the mass loss rate and mixing length of convection. Time variation of superficial abundances due to mass loss was first considered qualitatively by Michaud \& Charland (1986). They used a kinematic method and have shown that mass loss can produce variations of superficial abundances. But they have not solved the time-dependent continuity equation as in  Alecian (1996). Both of 
these models do not take into account the structural changes due
to diffusion as those of Richard, Michaud \& Richer (2001) for instance.

\emph{Scenario B} supposes that the surface abundances come from much deeper in
the star, near where Sc and Ca are in the Ne-like configuration (Richard,
Michaud \& Richer 2001, Richer, Michaud \& Turcotte 2000). A convection zone
created by an accumulation of iron in the region where $\log\;T \simeq$ 5.3  in
these evolutionary models including atomic diffusion (but while neglecting mass
loss) is at the base of this conclusion. The $g_{rad}$ of Fe shown in
Fig.~\ref{fig:gradFe} are consistent with an accumulation of this element in
this region of the star. However, as previously mentioned, mass loss could
modify the stratification profiles of Fe and of the other elements.

More recently (Michaud, Richer \& Richard 2005), these models were applied to
the star $o$ Leo A, and lead to similar abundances as observed by Griffin (2002)
for this star. These authors mentioned that this star serves as a severe test
for the deep mixing framework (\emph{scenario B}). They also mention that a
model supposing surface abundances created below the superficial hydrogen
convection zone (Watson 1971) would be diluted during the subgiant phase. This
point will be discussed in the conclusion. They also state that other species
that could be compared to observations would be useful.

In Figure~\ref{fig:gradSc}, we notice that in the region of the Ne-like
configurations for Ca and Sc and where the abundance anomalies occur in
\emph{scenario B} (at $\log\,T \simeq 5.3$), the $g_{rad}$ of Sc are larger than
those of Ca, and even larger than gravity up to $\log\,T \simeq 5.6$. It is
instructive to look at the curves with an abundance of 0.1 times solar values in
Fig.~\ref{fig:gradSc}, since in Am stars, both Ca and Sc are generally
underabundant. These curves show that if \emph{scenario B} is the correct one,
the mixing has to be deeper than $\log\,T = 5.7$ for Sc to be more underabundant
than Ca (relative to solar values). This is deeper than the mixing predicted in
the models of Richard, Michaud \& Richer (2001).

In the stellar model presented here, the bottom of the hydrogen convection zone
is located at $\log\,T = 4.3$. This is not deep enough to explain the relative
underabundances of Ca and Sc simultaneously, if one strictly considers the
original model of Watson (1971). But, this is not incompatible with
\emph{Scenario A}. The region where the $g_{rad}$ of Sc is weaker than those of
Ca and that is consistent with their observed relative abundances is located
near $\log\,T = 4.6$ for abundances of 0.1 times solar values. In this 
region, $g_{rad}$ of both Ca and Sc are weaker than gravity for solar abundances.
According to numerical modelling of a star
not very far from the present situation, Alecian (1996) showed that, assuming
mass loss and considering that the bottom of the mixing zone is uncertain,
\emph{Scenario A} could account for such simultaneous underabundances.
However, only detailed evolutionary calculations
while including diffusion will be able to determine which of the two
scenarios is correct.

\section{Conclusion} \label{sec:conclusion}

The radiative accelerations of Sc presented here shed new light on the source of
Am stars' abundance anomalies and represents a new diagnostic tool to test
various theoretical models. Scandium could to be a key element to better
understanding these stars.

Our results rekindle the question as to where the abundance anomalies emanate
from in Am stars: either from a superficial or a much deeper mixing zone. The
results shown here regarding $g_{rad}$ of Sc suggest that for \emph{scenario B}
to be able to reproduce both the observed Ca and Sc abundances, the mixing must
be deeper than that found in the models of Talon, Richard \& Michaud (2006) or
those of Richard, Michaud \& Richer (2001) for instance. In these models, the
mixing goes down to approximately $\log\,T = 5.3$, while from the results shown
here, it has to go down to approximately $\log\,T = 5.7$ to be able to reproduce
both Ca and Sc observed abundances.

Relating to the question of dilution of the abundances during the subgiant phase
of evolution, it is not clear if this argument can eliminate \emph{scenario A}.
Even though the surface underabundances are created at much shallower regions of
the star in this scenario compared to \emph{scenario B}, there should also exist
an underabundance of both Ca and Sc in the two regions where the noble gas
configurations dominate. The value of the overabundance that should prevail
between these two regions as well as the depth of the mixing zone will determine
to what extent the surface abundances are affected at the subgiant stage. Only
detailed evolutionary calculations while including the deeper mixing at the
subgiant stage could show how the surface abundances of \emph{scenario A} are
affected at that stage of evolution. The predicted abundances of other elements
should also give insight into the amplitude of the dilution.

Our results show that near $\log\,T = 4.6$ (for abundances of 0.1 times solar
values), the $g_{rad}$ of Ca are stronger than those of Sc. \emph{Scenario A}
could then produce Ca and Sc abundances consistent with their relative observed
abundances in Am stars. The results presented here for Sc thus reinforce
the superficial mixing zone scenario but only evolutionary stellar models
with mass loss and diffusion, while including Sc, might help to finally solve
the Am stars' mysteries. We hope that the $g_{rad}$ of Sc shown here, or others
calculated directly from reliable atomic data, will be included in evolutionary
models in the near future. For that purpose, the parameters for Sc
calculated here are made available at http://www.umoncton.ca/leblanfn/grad.

\begin{acknowledgements} Authors thank O.~Richard for kindly communicating the
stellar model used in this paper. This research was partially funded by NSERC
and La Facult\'e des \'Etudes Sup\'erieures et de la Recherche de l'Universit\'e
de Moncton. We also thank RQCHP for computing time. One of us (FL) is grateful
for a one month visiting position at l'Observatoire de Meudon and at
l'Universit\'e de Paris 7. We acknowledge the financial support of Programme
National de Physique Stellaire (PNPS) of CNRS/INSU, France.
\end{acknowledgements}


\begin{thebibliography}{}


\bibitem[\protect\citeauthoryear{Abt}{1961}]{abt61}Abt, H. A. 1961, ApJS, 6, 37

\bibitem[\protect\citeauthoryear{Adelman}{1994}]{ade94} Adelman, S. J. 1994,
MNRAS, 271, 355

\bibitem[\protect\citeauthoryear{Alecian}{1985}]{ale85} Alecian, G. 1985, A\&A,
145, 275

\bibitem[\protect\citeauthoryear{Alecian}{1986}]{ale86} Alecian, G. 1986, A\&A,
168, 204

\bibitem[\protect\citeauthoryear{AlecianArtrua}{1990}]{ale90a} Alecian, G. \&
Artru, M.-C. 1990a, A\&A, 234, 323

\bibitem[\protect\citeauthoryear{AlecianArtrub}{1990}]{ale90b} Alecian, G. \&
Artru, M.-C. 1990b, A\&AS, 83, 379

\bibitem[\protect\citeauthoryear{Alecian}{1994}]{ale94} Alecian, G. 1994, A\&A,
289, 885

\bibitem[\protect\citeauthoryear{Alecian}{1996}]{ale96} Alecian, G. 1996, A\&A,
310, 872

\bibitem[\protect\citeauthoryear{AlecianLeBlanc}{2000}]{ale00} Alecian, G. \&
LeBlanc, F. 2000, MNRAS, 319, 677

\bibitem[\protect\citeauthoryear{AlecianLeBlanc}{2002}]{ale02} Alecian, G. \&
LeBlanc, F. 2002, MNRAS, 332, 891 (Paper I)

\bibitem[\protect\citeauthoryear{Griffin}{2002}]{gri02} Griffin, R. E. 2002,
AJ, 123, 988

\bibitem[\protect\citeauthoryear{Hui-Bon-Hoa et al.}{1997}]{hui97} Hui-Bon-Hoa,
A., Burkhart, \& Alecian, G. 1997, A\&A, 323, 901

\bibitem[\protect\citeauthoryear{Hui-Bon-HoaAlecian}{1998}]{hui98} Hui-Bon-Hoa,
A. \& Alecian, G. 1998, A\&A, 332, 224

\bibitem[\protect\citeauthoryear{Hui-Bon-Hoa}{1999}]{hui99} Hui-Bon-Hoa, A.
1999, A\&A, 343, 261

\bibitem[\protect\citeauthoryear{Kurucz}{1990}]{kur90} Kurucz, R. L. 1990,
Trans. IAU, Vol. XXB, ed. McNally, D., (Kluwer), 168

\bibitem[\protect\citeauthoryear{KuruczPeytremann}{1975}]{kur75} Kurucz, R. L.
\& Peytremann, E. 1975, Smithsonian Astrophys. Obs. Spce. Rep., 362

\bibitem[\protect\citeauthoryear{LeBlanc et al.}{2000}]{leb00} LeBlanc, F.,
Michaud, G. \& Richer, J. 2000, ApJ, 538, 876

\bibitem[\protect\citeauthoryear{LeBlancAlecian}{2004}]{leb04} LeBlanc, F. \&
Alecian, G. 2004, MNRAS, 352, 1329 (Paper II)

\bibitem[\protect\citeauthoryear{Michaud}{1970}]{mic70} Michaud, G. 1970, ApJ,
160, 641

\bibitem[\protect\citeauthoryear{Michaud et al.}{1983}]{mic83} Michaud, G.,
Tarasick, D., Charland, Y. \& Pelletier, C. 1983, ApJ, 269, 239

\bibitem[\protect\citeauthoryear{MichaudCharland}{1986}]{mic86} Michaud, G.
\& Charland, Y. 1986, ApJ, 311, 326

\bibitem[\protect\citeauthoryear{Michaud et al.}{2005}]{mic05} Michaud, G.,
Richer, J. \& Richard, O. 2005, ApJ, 623, 442

\bibitem[\protect\citeauthoryear{Preston}{1974}]{pre74} Preston, G. W. 1974,
ARA\&A, 12, 257

\bibitem[\protect\citeauthoryear{RogersIglesias}{1992}]{rog92} Rogers, F. J. \&
Iglesias, C. A. 1992, ApJ, 401, 361

\bibitem[\protect\citeauthoryear{Richard et al.}{2001}]{ric01} Richard, O.,
Michaud, G. \& Richer, J. 2001, ApJ, 558, 377

\bibitem[\protect\citeauthoryear{Richer et al.}{2000}]{ric00} Richer, J.,
Michaud, G. \& Turcotte, S. 2000, ApJ, 529, 338

\bibitem[\protect\citeauthoryear{Seaton}{1997}]{sea97} Seaton, M. J. 1997,
MNRAS, 289, 700

\bibitem[\protect\citeauthoryear{Seaton et al.}{1992}]{sea92} Seaton, M. J.,
Zeippen, C. J., Tully, J. A., Pradham, A. K., Mendoza, C., Hibbert, A. \&
Berrington, K. A. 1992, Rev. Mex. Astron. Astrofis., 23, 19

\bibitem[\protect\citeauthoryear{Smith}{1971}]{smi71} Smith, M. A. 1971, A\&A,
11, 325

\bibitem[\protect\citeauthoryear{Talon et al.}{2006}]{tal06} Talon S., Richard,
O. \& Michaud, G. 2006, ApJ, 645, 634

\bibitem[\protect\citeauthoryear{Watson}{1971}]{wat71} Watson, W. D. 1971, A\&A,
13, 263


\end{thebibliography}
\end{document}